# The EU-U.S. Data Privacy Framework: Is the Dragon Eating its Own Tail?

Marcelo Corrales Compagnucci∗

**Abstract:** The European Commission's adequacy decision on the EU-U.S. Data Privacy Framework (DPF), adopted on July 10th, 2023, marks a crucial moment in transatlantic data protection. Following an Executive Order issued by President Biden in October 2022, this decision confirms that the United States (U.S.) meets European Union (EU) standards for personal data protection. The decision extends to all transfers from the European Economic Area (EEA) to U.S. entities participating in the framework, promoting privacy rights while facilitating data exchange. Key aspects include oversight of U.S. public authorities' access to transferred data, the introduction of a dual-tier redress mechanism, and granting new rights to EU individuals, encompassing data access and rectification. However, the EU-U.S. DPF presents both promise and challenges in health data transfers. While streamlining exchange and aligning legal standards, it grapples with the complexities of divergent privacy laws. The recent bill for the introduction of a U.S. federal privacy law emphasizes the urgent need for ongoing reform. Lingering concerns persist regarding the EU-U.S. DPF's resilience, especially amid potential legal battles before the Court of Justice of the EU (CJEU). The history of transatlantic data transfers between the EU and the U.S. is riddled with vulnerabilities, reminiscent of the Ouroboros – an ancient symbol of a serpent or dragon eating its own tail – hinting at the looming possibility of the framework facing invalidation once again. This article delves into the main requirements of the EU-U.S. DPF and offers insights on how healthcare organizations can navigate it effectively.

---

# 1 Introduction

In ancient mythology, there exists a symbol known as the Ouroboros – a serpent or dragon devouring its own tail, forming a perfect circle. This symbolizes the eternal cycle of life, death, and rebirth, as well as the concept of self-renewal and infinity. The Ouroboros embodies the idea of interconnectedness and the cyclical nature of existence.[1]

In the realm of international data transfer mechanisms and the history of data privacy frameworks between the EU and the U.S., we can draw a similar analogy to the Ouroboros. Just as the Ouroboros represents a cycle of perpetual renewal, the landscape of data protection and privacy law undergoes a similar process of constant evolution and transformation. This analogy encapsulates the European Commission's struggle to maintain privacy standards amidst legal challenges to data transfer mechanisms. It underscores the cyclical nature of these challenges, where efforts to protect privacy are met with scrutiny and the need for continual adaptation and innovation.

Initially, mechanisms like the Safe Harbor and Privacy Shield programs were established. However, as time passed and concerns about the adequacy of data protection grew, the European Commission and the Court of Justice of the EU (CJEU) found themselves confronting legal challenges and criticisms, particularly in cases like *Schrems I*[2] and *Schrems II*[3] which revealed gaps in the protection of personal data transferred from the EU to the U.S. Concerns about U.S. surveillance practices and insufficient safeguards led to the invalidation of both previous frameworks, leading the new EU-U.S. Data Privacy Framework (DPF).[4]

The EU-U.S. DPF offers a structured approach for companies in the U.S., including healthcare organizations to ensure compliance with General Data Protection Regulation (GDPR)[5] when handling cross-border personal health data transfers from the EU. Participating in the EU-U.S. DPF provides advantages such as simplified legal compliance, streamlined collaboration with EU healthcare entities, improved access to international markets, and enhanced trust among patients and partners. However, involvement in the EU-U.S. DPF subjects healthcare organizations to oversight by the U.S. Federal Trade Commission (FTC) and requires thorough due diligence efforts. Moreover, there's a potential risk of the framework being legally challenged, akin to past frameworks, necessitating careful strategic decision-making within the healthcare sector. Overall, while offering benefits, participation in the framework demands thoughtful consideration of its implications for healthcare organizations and their patients' privacy rights.

---

[1] McCullen (2021), p. 40.
[2] C-362/14 Maximiliam Schrems v Data Protection Commissioner of 6 October 2016 [2015] ECLI:EU:C:2015:650 (*Schrems I*). See also, CJEU Press Release No. 117/15 (6 October 2015).
[3] C-311/18 Data Protection Commissioner v. Facebook Ireland Limited, Maxilimian Schrems (*Schrems II*) [2020] ECLI: EU:C:2020:559
[4] See Implementing Decision of 10.7.2023 pursuant to Regulation (EU) 2016/679 of the European Parliament and of the Council on the adequate level of protection of personal data under the EU-US Data Privacy Framework. Available at: https://commission.europa.eu/system/files/2023-07/Adequacy%20decision%20EU-US%20Data%20Privacy%20Framework_en.pdf (Accessed 14 April 2024).
[5] Regulation (EU) 2016/679 of the EP and of the Council of 27 April 2016 on the protection of natural persons with regard to the processing of personal data and on the free movement of such data, and repealing Directive 95/46/EC, OJ 2016 L 119, 1 (General Data Protection Regulation, GDPR).

After this introduction, this article is structured as follows. Section 2 discusses how the GDPR impacts data transfers, particularly in healthcare using cloud computing and other new technologies. It highlights challenges like cloud computing's borderless nature and complexities in transferring sensitive healthcare data while ensuring compliance. Section 3 explains the *Schrems I & II* cases before the CJEU, uncovering the historical shortcomings of the previous Safe Harbor and Privacy Shield frameworks. Section 4 delves into the details of the EU-U.S. DPF, shedding light on the main principles, including its novel dual-tier redress mechanism. Section 5 explores the opportunities and challenges presented by this new framework for the healthcare sector. Finally, section 6 concludes.

## 2 Cloud Computing, AI and GDPR: Implications for Health Data Transfers

The enactment of the GDPR has catalyzed significant responses globally, prompting scrutiny from EU institutions and nations worldwide. While the GDPR has largely been lauded for rectifying shortcomings of the preceding Data Protection Directive,[6] the currently regulatory framework still grapples with unresolved issues. Of particular concern is the adequacy of the proposed reform in accommodating the evolving landscape of new technologies and the healthcare sector, notably the utilization of cloud computing for the transfer and sharing of patient data across borders, such as in clinical trials. As more healthcare organizations transition to cloud-based solutions, they unlock numerous advantages such as scalability, flexibility, and cost-efficiency. However, this shift also introduced several challenges, particularly regarding data privacy.[7]

One consequence of cloud computing is its inherent location agnosticism.[8] The ubiquitous architecture of cloud systems allows for the dynamic provisioning of data and resources, rendering obsolete the traditional boundaries upon which many data protection regulations were established. For example, datasets can swiftly traverse jurisdictions within seconds, contingent upon the virtual space available on a server. This reality poses significant challenges to current data protection laws, which often prioritize data location and typically prohibit the transfer of personal data to third countries without legal exceptions. These exceptions traditionally assume data exists in physically tangible forms, such as paper files or stored devices, necessitating physical transfers between locations. However, in the cloud environment, this paradigm does not always apply. Recognizing these challenges, the GDPR has sought to address this issue by updating the provisions that govern the transfer of personal data outside the EU, ensuring that data protection standards remain robust even in the context of cloud computing.[9]

Cloud computing, combined with mobile computing and sensors, has revolutionized the provision and management of healthcare services. This transformation has been further accelerated by the increasing utilization of wearable devices, exacerbating the challenges associated with data privacy, security, and compliance. This concern stems from their intrinsic volatile nature, which undermines current data protection and security mechanisms. Several solutions have emerged to address challenges at the intersection of healthcare, cloud computing and wearable devices. Moreover, there is a growing emphasis on integrating artificial intelligence (AI)[10] with cloud technology for various medical applications, underscoring the ongoing necessity for enhancing security protocols to effectively mitigate emerging threats.[11]

---

[6] Directive 95/46/EC of the European Parliament and of the Council of 24 October 1995 on the protection of individuals with regard to the processing of personal data and on the free movement of such data, OJ 1995 L 281, 355 ('EU Data Protection Directive').
[7] Kirkham et al. (2012), pp. 1063 et seq; Barnitzke, Corrales Compagnucci and Forgó (2012); Corrales Compagnucci, Kousiouris and Vafiadis (2013), pp. 61-72.
[8] Sharma and Sood (2014), p. 652.
[9] See Chapter V of the GDPR.

These challenges encompass a spectrum including data availability, integrity, confidentiality, and network security.[12] Proposed solutions include a range of measures such as data multiparty encryption, authentication protocols, data classification frameworks, and the implementation of application programming interfaces (APIs).[13] The paramount importance of preserving patient data privacy and confidentiality while harnessing the advantages of new technologies in healthcare cannot be overstated. It is essential to ensure that patient sensitive data remains safeguarded against unauthorized access or breaches, maintaining the trust and integrity of the healthcare system.[14]

Consider a scenario where a healthcare provider implements robust encryption protocols within cloud-based electronic health records (EHR) systems.[15] These systems not only securely store patient data but also allow authorized medical professionals to efficiently access and update records as needed, enhancing collaboration and patient care while maintaining strict adherence to privacy regulations and ethical standards. In the realm of cloud computing, healthcare organizations often entrust sensitive information to third-party providers, raising concerns about data control and security. For example, data stored in shared environments could be at risk of unauthorized access or leakage, posing a threat to confidentiality.

Additionally, determining the applicable legal framework becomes complex due to the cloud's decentralized nature, where data may reside in multiple jurisdictions simultaneously. One recurring concern revolves around maintaining privacy commitments across different jurisdictions. Organizations expect cloud service providers to uphold privacy standards consistent with their own, yet variations in regulations across regions complicate this expectation. Moreover, GDPR-specific challenges further compound the complexities of cloud and AI adoption.[16]

In sum, the transfer of sensitive data to third countries, particularly those lacking robust data protection measures, poses a significant challenge. Despite the advancements introduced by the GDPR in refining mechanisms from previous regimes, the current legal landscape remains intricate and challenging to navigate. In addressing these complexities, the following sections aim to provide clarity on the current legal framework, in particular for data transfers between the EU and the U.S. within the new framework. By examining the evolution of data transfer practices across the Atlantic, the following sections aim to equip stakeholders in the healthcare sector with the knowledge and resources necessary to navigate this complex terrain. Through informed decision-making and proactive measures, the goal is to facilitate the secure and compliant transfer of sensitive healthcare data across borders, ensuring adherence to stringent data protection standards while leveraging the potential of new technologies to advance healthcare delivery.

---

[10] Panagopoulos et al. (2022).
[11] Kiran et al. (2013).
[12] Barnitzke (2011); Corrales Compagnucci (2019), pp. 144-155.
[13] Minssen et al. (2020), pp. 34-50.
[14] Mehrtak et al. (2021), pp. 448-461.
[15] Ahmadi and Aslani (2018), pp. 24-28.
[16] Corrales Compagnucci (2019).

## 3 *Schrems I & II*: Unraveling the Shortcomings of the Safe Harbor and Privacy Shield Previous Frameworks

The Safe Harbor program for personal data transfers was a framework established by the U.S. Department of Commerce in cooperation with the European Commission in 2000. It was the first framework designed to provide a method for companies in the U.S. to comply with the EU's Directive on Data Protection, which prohibited the transfer of personal data to non-EU countries that did not meet the EU's adequacy standards for data protection.[17]

Under the Safe Harbor program, participating companies in the U.S. could self-certify their compliance with certain privacy principles and requirements deemed adequate by the European Commission. These principles covered aspects such as notice, choice, onward transfer, security, data integrity, access, and enforcement. By adhering to these principles, U.S. organizations could lawfully receive personal data from the EU without additional authorization being required for each transfer.[18]

Despite the promising compromise that the Safe Harbor program offered for transatlantic data transfer, numerous critical findings and legal challenges emerged, revealing the limitations within this self-regulatory and voluntary mechanism. For example, the Galexia study assessed each of the organizations listed on the Safe Harbor List against a limited subset of key criteria outlined in the Safe Harbor framework principles. The assessment revealed numerous instances of data inaccuracy and failure to comply with basic requirements of the framework. Specifically focusing on Principle 7 – Enforcement and Dispute Resolution, where only 348 organizations out of 1,597 were found to pass this basic compliance test. Overall, the study concluded that the issues identified in previous reviews of the Safe Harbor remained unaddressed, with the prevalence of false claims made by organizations posing a significant privacy risk to consumers.[19]

Additionally, the Article 29 Working Party suggested that Safe Harbor, along with other adequacy findings, had limitations concerning their geographical scope, leaving certain transfers within cloud computing transformations uncovered. While transfers to U.S. organizations adhering to Safe Harbor principles were lawful under EU law, the Working Party expressed concerns regarding the sole reliance on Safe Harbor self-certification without robust enforcement of data protection principles in the cloud environment. Furthermore, the Working Party emphasized the importance of contracts between controllers and processors for processing purposes, as outlined in Article 17 of the EU Data Protection Directive and FAQ 10 of the EU-U.S. Safe Harbor framework documents. It was noted that such contracts should specify the processing activities and security measures, although they did not require prior authorization from European Data Protection Authorities (DPAs).[20]

---

[17] Westboy (2004), p. 95.
[18] International Trade Administration, Department of Congress, Safe Harbor Overview, (2002), p. 619.
[19] Connolly (2008).

The Working Party recommended that data-exporting organizations should not solely depend on the statement of the data importer's Safe Harbor certification. Instead, they should obtain evidence of the certification's existence and compliance with its principles, particularly concerning information provided to affected data subjects. Additionally, the Working Party advised cloud users to ensure that standard contracts offered by cloud providers comply with national requirements for contractual data processing. In cases where cloud providers did not provide sufficient information, exporters were encouraged to utilize other legal instruments such as Standard Contractual Clauses (SCCs) or Binding Corporate Rules (BCR).[21]

Overall, the Working Party cautioned that adherence to Safe Harbor principles alone might not guarantee adequate security measures by U.S. cloud providers, as required by national legislations based on Directive 95/46/EC. Cloud computing posed specific security risks not adequately addressed by existing Safe Harbor principles, necessitating additional safeguards such as third-party assessments, standardization, and certification schemes tailored to the cloud environment. Therefore, complementing Safe Harbor commitments with additional safeguards specific to cloud computing was deemed advisable.[22]

A major concern arose regarding the potential unauthorized access to personal data, prompting fears of infringements upon the originally intended purposes for which such data was transferred. A significant portion of major U.S. Internet companies, allegedly involved in surveillance programs, held certifications under the Safe Harbor. Given these identified deficiencies, the implementation of the framework was deemed unsustainable, prompting an urgent call for reassessment and reform. Complicating matters further, the Safe Harbor framework heavily relied on the "commitments" and "self-certification" of participating companies, albeit with legally binding regulations.[23]

The legal discourse surrounding international data transfers intertwines with two subsequent cases adjudicated by the CJEU. In the first case, Austrian privacy activist named Max Schrems lodged a complaint in June 2013 with the Irish Data Protection Commission (DPC) (*Schrems I*).[24] His complaint urged the supervisory authority to intervene and cease the transfer of his personal data from Facebook Ireland to the U.S. Schrems broadly argued that the legal framework and practices in the U.S. – such as section 702 of the Foreign Intelligence Surveillance Act (FISA) and Executive Order 12333 – failed to ensure adequate protection for personal data against surveillance activities carried out by public authorities.[25]

Specifically, he was concerned about the surveillance activities conducted by various public authorities, including the U.S. National Security Agency (NSA). U.S. surveillance laws and programs revealed by the Snowden disclosures,[26] such as PRISM,[27] allowed U.S. authorities unrestricted access to personal data collected by U.S. Big Tech companies, including Facebook. The CJEU upheld Schrems' arguments and, in 2015, overturned the Safe Harbour Agreement,[28] deeming the transfer of personal data illegal under this framework. The CJEU ruled that this action constituted a violation of European privacy laws and the fundamental human rights principles enshrined in Articles 7, 8, and 47 of the Charter of Fundamental Rights (CFR) of the EU.[29] After the Safe Harbor Agreement was invalidated, Facebook and other companies turned to SCCs as a solution. SCCs are among the most commonly used mechanisms authorized by the European Commission for transferring personal data outside the EU.[30]

---

[20] Article 29 Data Protection Working Party, Opinion 05/2012 on Cloud Computing, adopted 1 July 2012, p. 17.
[21] Article 29 Data Protection Working Party, Opinion 05/2012 on Cloud Computing, adopted 1 July 2012, p. 17-18.
[22] Article 29 Data Protection Working Party, Opinion 05/2012 on Cloud Computing, adopted 1 July 2012, p. 17-18.
[23] Article 29 Data Protection Working Party, Opinion 05/2012 on Cloud Computing, adopted 1 July 2012, p. 17-18.
[24] C-362/14, Maximillian Schrems vs Data Protection Commissioner (*Schrems I*).
[25] Jurcys, Corrales Compagnucci and Fenwick (2022).

In response to the *Schrems I* decision and the subsequent invalidation of the Safe Harbor framework, the EU Commission introduced a new data transfer tool in July 2016 known as the EU-U.S. Privacy Shield.[31] This framework was built on four key pillars:[32]

1. Transparency: Companies engaged in data transfers were mandated to be transparent about their practices.
2. Enhanced Limitations: Stricter limitations were imposed on companies, along with the development of more adequate supervision mechanisms.
3. Legal Redress: Increased opportunities for legal redress and more effective mechanisms for alternative dispute resolution (ADR), including the implementation of an arbitration procedure.
4. Annual Review: The framework included provisions for an annual review involving both EU and U.S. authorities, as well as other stakeholders.

Since the U.S. did not fall under the category of countries considered to have an adequate level of data protection in the assessment of the EU, the EU-U.S. Privacy Shield program introduced a set of requirements, including the submission of a privacy policy with specific details. It emphasized greater transparency, oversight, and redress mechanisms, which encompassed the establishment of an ombudsman to investigate complaints. The Privacy Shield promised to effectively facilitate cross-border transfers from the EU to the U.S. by operating within the framework of Article 45 of the GDPR, serving as a limited adequacy decision.[33] This framework provided a legitimate pathway for transferring personal data to U.S.

---

[26] Gellman (2020).
[27] Watt (2013).
[28] 2000/520/EC: Commission Decision of 26 July 2000 pursuant to Directive 95/46/EC of the European Parliament and of the Council on the adequacy of the protection provided by the safe harbour privacy principles and related frequently asked questions issued by the US Department of Commerce (notified under document number C(2000) 2441).
[29] Corrales Compagnucci et al. (2020), pp. 154-155.
[30] Jurcys, Corrales Compagnucci and Fenwick (2022). For more information about SCCs, see Corrales Compagnucci, Aboy and Minssen (2021), pp. 37-47.
[31] Commission Implementing Decision (EU) 2016/1250 of 12 July 2016 pursuant to Directive 95/46/EC of the European Parliament and of the Council on the adequacy of the protection provided by the EU-US Privacy Shield (notified under document C(2016) 4176).
[32] Jurcys, Corrales Compagnucci and Fenwick (2022).
[33] Bradford, Aboy and Liddell (2020).

organizations that could demonstrate they offered adequate protection for data transferred from the EU.

However, in December 2015, Max Schrems submitted an amended complaint to the Irish Data Protection Commissioner (DPC), contesting the legitimacy of Facebook's utilization of SCCs. He urged the DPC to either forbid or temporarily halt the transmission of his personal data to Facebook Inc., arguing that the SCCs were insufficient to safeguard against potential access by US authorities, particularly considering revelations about mass surveillance practices by U.S. intelligence agencies, following the Snowden disclosures.[34]

This complaint ultimately led to the landmark *Schrems II* decision. In July 2020, the CJEU, in the case C-311/18 *Data Protection Commissioner v Facebook Ireland Limited, Maximillian Schrems*,[35] declared the EU-U.S. Privacy Shield null and void, while confirming the validity of SCCs.[36] The Court stipulated a "case-by-case" assessment regarding the application of SCCs. Controllers[37] and processors[38] exporting data are obligated to ascertain whether the legislation and practices of the third country undermine the efficacy of the appropriate safeguards outlined in Article 46 of the GDPR. The *Schrems II* ruling implies that data exporters must implement "supplementary measures" to address any deficiencies and ensure compliance with EU law. Regrettably, the CJEU did not provide a clear definition or specification of these "supplementary measures". This ambiguity sparked intense debates and prompted the issuance of numerous guidelines and recommendations on the implementation of additional safeguards.[39]

In its examination of the Privacy Shield framework, the CJEU invalidated it on two primary grounds. Firstly, the Court found that the U.S. surveillance programs, as evaluated by the Commission in its decision on the Privacy Shield, failed to comply with the stringent requirements of EU law. Specifically, these programs were deemed neither strictly necessary nor proportionate, thus contravening Article 52 of the EU CFR. Secondly, the Court determined that, with regard to U.S. surveillance practices, EU data subjects lacked accessible and effective judicial remedies, depriving them of the right to redress in the U.S., as mandated by Article 47 of the EU CFR.[40]

---

[34] See, Explanatory Memorandum on the Litigation concerning Standard Contractual Clauses (SCCs), Irish Data Protection Commission, available at: https://www.dataprotection.ie/en/dpc-guidance/law/scclitigation (Accessed 27 March 2024).
[35] C-311/18 Data Protection Commissioner v. Facebook Ireland Limited, Maxilimian Schrems (*Schrems II*) [2020] ECLI: EU:C:2020:559.
[36] Corrales Compagnucci, Aboy and Minssen (2021), p. 40.
[37] Within the framework of the GDPR, a "controller" is the entity entrusted with determining "the means and purposes of the processing" (Article 4(7) GDPR).
[38] A "processor" is the entity engaged in processing personal data on behalf of the controller, typically through outsourcing arrangements (Article 4(8) GDPR).
[39] See, e.g., European Commission implementing decision of 4 June 2021 on standard contractual clauses for the transfer of personal data to third countries pursuant to Regulation (EU) 2016/679 of the European Parliament and of the Council, C(2021) 3972 final; EDPB Recommendations on measures that supplement transfer tools to ensure compliance with the EU level of protection of personal data. Adopted on 10 November 2020, available at: https://www.edpb.europa.eu/our-work-tools/our-documents/recommendations/recommendations-012020-measures-supplement-transfer_en (Accessed 27 March 2024); Corrales Compagnucci, Aboy and Minssen (2021), pp. 40-41.
[40] Caitlin Fennessy (2020).

# 4 The EU-U.S. Data Privacy Framework (DPF) – Third Time Lucky?

In July 2023, a significant milestone was reached within the landscape of transatlantic data protection as the European Commission officially adopted its third adequacy decision concerning the EU-US Data Privacy Framework. This decision affirmed that the U.S. guarantees an appropriate level of protection for personal data transferred from the European Union, aligning with the standards upheld within the EU.[41]

The adequacy decision followed the issuance of an Executive Order (EO 14086)[42] titled "Enhancing Safeguards for United States Signals Intelligence Activities", endorsed by President Biden on October 7th, and complemented by regulations sanctioned by the U.S. Attorney General.[43] This executive action introduced stringent measures aimed at mitigating concerns highlighted in the *Schrems II* decision of July 2020. Significantly, these new obligations were designed to restrict access to data by U.S. intelligence agencies strictly to what is deemed necessary and proportionate. Additionally, they sought to institute an independent and impartial redress mechanism to address and adjudicate complaints from European individuals pertaining to the gathering of their data for national security purposes.[44]

The adequacy decision concerning the EU-U.S. DPF extends its purview to encompass data transfers from any entity, whether public or private, within the European Economic Area (EEA) to U.S. companies actively engaged in the EU-U.S. DPF. This agreement serves as a cornerstone in fostering transatlantic data flows, ensuring the preservation of privacy rights while facilitating the exchange of information between the EU and the U.S.[45]

Within this adequacy decision, the Commission scrutinized the requirements outlined in the EU-U.S. DPF, including the limitations and safeguards concerning access to transferred personal data by U.S. public authorities, notably within criminal law enforcement and national security purposes. Based on this careful assessment, the adequacy decision asserts the U.S.'s provision of a sufficient level of protection for personal data transferred from the EU to entities participating in the EU-U.S. DPF. With the formal adoption of this decision, European entities can transfer personal data to U.S. counterparts engaged in the framework without the need for additional data protection safeguards.[46]

The framework grants EU individuals, whose data may be transferred to participating U.S. companies, with a range of new rights. These include the right to access their data, rectify inaccuracies, or request deletion of data that is incorrect or unlawfully processed. Moreover, it establishes various redress mechanisms in the event of mishandling of data, encompassing free-of-charge independent dispute resolution mechanisms and an arbitration panel.[47]

---

[41] Questions and answers EU-US Data Privacy Framework (10 July 2023), available at: https://ec.europa.eu/commission/presscorner/detail/en/qanda_23_3752 (Accessed 27 March 2024).
[42] Executive Order 14086 – Policy and Procedures. Bureau of Intelligence and Research, available at: https://www.state.gov/executive-order-14086-policy-and-procedures/ (Accessed 27 March 2024).
[43] The Data Protection Review Court, Office of Privacy and Civil Liberties, U.S. Department of Justice, available at: https://www.justice.gov/opcl/redress-data-protection-review-court (Accessed 27 March 2024).
[44] European Commission and United States Joint Statement on Trans-Atlantic Data Privacy Framework, Press Release (25 March 2022), available at: https://ec.europa.eu/commission/presscorner/detail/en/ip_22_2087 (Accessed 27 March 2024).
[45] European Commission and United States Joint Statement on Trans-Atlantic Data Privacy Framework, Press Release (25 March 2022), available at: https://ec.europa.eu/commission/presscorner/detail/en/ip_22_2087 (Accessed 27 March 2024).
[46] European Commission and United States Joint Statement on Trans-Atlantic Data Privacy Framework, Press Release (25 March 2022), available at: https://ec.europa.eu/commission/presscorner/detail/en/ip_22_2087 (Accessed 27 March 2024).

Similarly to the previous Privacy Shield Framework, U.S. organizations can become certified participants in the EU-U.S. DPF by pledging to adhere to a detailed array of privacy obligations. This commitment encompasses adherence to privacy principles such as purpose limitation, data minimization, and data retention, alongside specific responsibilities pertaining to data security and the sharing of data with third parties.[48]

The administration of the framework falls under the responsibility of the U.S. Department of Commerce, tasked with processing certification applications and ensuring ongoing compliance of participating companies with the established requirements. Oversight of U.S. companies' adherence to their obligations within the EU-US DPF will be enforced by the U.S. FTC.

## 4.1 Limitations and Safeguards Regarding Access to Data by U.S. Intelligence Agencies

An integral element of the U.S. legal framework underpinning the adequacy decision centers around EO 14086, which is further reinforced by regulations authorized by the Attorney General. These measures were implemented to tackle the concerns highlighted by the CJEU in its *Schrems II* judgment.[49]

For individuals in Europe whose personal data is transferred to the U.S., EO 14086 presents a significant milestone. It introduces a set of robust safeguards designed to curtail the access of U.S. intelligence authorities to data, restricting it strictly to what is deemed necessary and proportionate for the preservation of national security interests. These safeguards are implemented to uphold the principles of privacy and civil liberties, aligning with the expectations set forth by European regulatory standards.[50]

Additionally, the EO amplifies the oversight mechanisms governing U.S. intelligence services. This heightened oversight attempts to ensure stringent adherence to prescribed limitations on surveillance activities, thereby bolstering transparency and accountability within the intelligence community. Furthermore, the EO establishes a new redress mechanism, a hallmark feature of the transatlantic data transfer landscape. Its establishment underscores a commitment to fair and equitable resolution of disputes, aiming at fostering trust and cooperation between EU and U.S. entities engaged in data transfers. The details of the new redress mechanism are further explained below.[51]

---

[47] European Commission and United States Joint Statement on Trans-Atlantic Data Privacy Framework, Press Release (25 March 2022), available at: https://ec.europa.eu/commission/presscorner/detail/en/ip_22_2087 (Accessed 27 March 2024).
[48] European Commission and United States Joint Statement on Trans-Atlantic Data Privacy Framework, Press Release (25 March 2022), available at: https://ec.europa.eu/commission/presscorner/detail/en/ip_22_2087 (Accessed 27 March 2024).
[49] See, EO 14086, available at: https://www.justice.gov/opcl/executive-order-14086 (Accessed 27 March 2024).
[50] The White House, 'Fact Sheet: President Biden Signs Executive Order to Implement the European Union-U.S. Data Privacy Framework' (7 October 2022), available at: https://www.whitehouse.gov/briefing-room/statements-releases/2022/10/07/fact-sheet-president-biden-signs-executive-order-to-implement-the-european-union-u-s-data-privacy-framework/ (Accessed 27 March 2024).

## *4.2 New Redress Procedure*

The U.S. Government has implemented a new dual-tier redress mechanism, equipped with independent and binding authority powers, to address and resolve complaints from individuals whose data has been transferred from the EEA to U.S.-based companies regarding the collection and utilization of their data by U.S. intelligence agencies.[52]

To render a complaint admissible, individuals are not required to prove that their data was in fact collected by U.S. intelligence agencies. They may submit their complaints to their national DPA, which will facilitate the proper transmission of the complaint and provide the individual with any additional information regarding the procedure, including its outcome. This approach ensures that individuals can seek recourse from an authority close to home, in their own language. Complaints will be forwarded to the U.S. by the European Data Protection Board (EDPB).[53]

In the first layer, EU individuals will be able to lodge a complaint with the designated Civil Liberties Protection Officer (CLPO) within the U.S. intelligence community. This officer is entrusted with ensuring that U.S. intelligence agencies adhere to privacy regulations and uphold fundamental rights. In other words, the CLPO will operate in a manner akin to a data protection officer specifically designated for the U.S. intelligence community. This initial layer will enable the intelligence community to implement essential corrective actions and support the operations of the newly established Data Protection Review Court (DPRC). This ensures that all complainants can seek recourse without any conditions and guarantees a fully independent review process.[54]

In the second layer, EU individuals would have the right to appeal the decision of the CLPO through the new DPRC.[55] The Court is composed of six members[56] external to the US Government, appointed based on specific qualifications, and independent from governmental influence. Court members can only be removed for justifiable reasons, such as criminal misconduct or incapacity, and are empowered to conduct thorough investigations into complaints lodged by EU individuals. The DPRC holds authority to compel intelligence agencies to provide relevant information and issue legally binding remedial decisions. For instance, should the DPRC determine that data gathering breaches the safeguards outlined in the EO, it possesses the authority to mandate the deletion of such data.[57]

---

[51] The White House, 'Fact Sheet: President Biden Signs Executive Order to Implement the European Union-U.S. Data Privacy Framework' (7 October 2022), available at: https://www.whitehouse.gov/briefing-room/statements-releases/2022/10/07/fact-sheet-president-biden-signs-executive-order-to-implement-the-european-union-u-s-data-privacy-framework/ (Accessed 27 March 2024).
[52] Questions and answers EU-US Data Privacy Framework (10 July 2023), available at: https://ec.europa.eu/commission/presscorner/detail/en/qanda_23_3752 (Accessed 27 March 2024).
[53] Questions and answers EU-US Data Privacy Framework (10 July 2023), available at: https://ec.europa.eu/commission/presscorner/detail/en/qanda_23_3752 (Accessed 27 March 2024).
[54] Peter Swire (2022).
[55] Attorney General Merrick B. Garland Announces Judges of the Data Protection Review Court. Press Release (14 November 2023), available at: https://www.justice.gov/opa/pr/attorney-general-merrick-b-garland-announces-judges-data-protection-review-court (Accessed 27 March 2024).
[56] The judges on the Court include: James E. Baker, Rajesh De, James X. Dempsey, Mary B. DeRosa, Thomas B. Griffith, Eric H. Holder Jr., David F. Levi, and Virginia A Seitz. For full bios and more information on the Data protection Review Court, see https://www.justice.gov/opcl/redress-data-protection-review-court (Accessed 27 March 2024).

In each instance, the Court will appoint a dedicated special advocate with relevant experience to assist its proceedings. This advocate will ensure the representation of the complainant's interests, while also furnishing the Court with in-depth understanding of the factual and legal dimensions of the case. Such a measure guarantees equitable representation for both parties, thereby reinforcing essential guarantees in terms of fair trial and due process. Once the CLPO or the DPRC concludes the investigation, the complainant will be promptly notified either that no violation of US law was detected, or that a violation was indeed found and remedied. Furthermore, at a subsequent stage, the complainant will receive notification when any information regarding the DPRC proceedings – such as the reasoned decision of the Court – becomes unrestricted from confidentiality requirements and is available for access.[58]

### 4.3 Principles Enshrined in the EU-U.S. DPF

The U.S. and the EU share a commitment to enhancing privacy protection, despite differing approaches. The DPF aims to facilitate data transfers while ensuring EU data subjects' protection. Organizations in the US must self-certify adherence to the following principles explained below. Non-compliance results in removal from the list. Organizations removed must continue safeguarding data or return it to the EU. Adherence to the principles may be limited under specific circumstances, such as court orders or national security requirements.[59]

**4.3.1 Notice:** An organization must inform individuals about its participation in the EU-U.S. DPF, the types of personal data collected, and its commitment to adhere to the principles. This includes disclosing the purposes for data collection, contact information for inquiries or complaints, third-party disclosures, access rights, dispute resolution mechanisms, and its liability in cases of onward transfers to third parties. The notice should be provided in clear and conspicuous language before using or disclosing personal information for purposes other than originally intended.[60]

**4.3.2 Choice:** Organizations must provide individuals with the option to opt out of their personal information being disclosed to third parties or used for purposes different from the original intent. This choice should be easily accessible and clearly presented to individuals. However, there are exceptions, such as when disclosure is to an agent acting on behalf of the organization, though a contractual agreement is still required. When handling sensitive information, which includes medical or health conditions, racial or ethnic origin, and other personal details, organizations must obtain explicit consent (opt-in) from individuals before disclosing or using such data for any purpose other than originally intended. This requirement applies even when the information is received from a third party who identifies it as sensitive.[61]

---

[57] Questions and answers EU-US Data Privacy Framework (10 July 2023), available at: https://ec.europa.eu/commission/presscorner/detail/en/qanda_23_3752 (Accessed 27 March 2024).
[58] Questions and answers EU-US Data Privacy Framework (10 July 2023), available at: https://ec.europa.eu/commission/presscorner/detail/en/qanda_23_3752 (Accessed 27 March 2024).
[59] Annex 1. EU-U.S. Data Privacy Framework Principles issued by the U.S. Department of Commerce, available at: https://commission.europa.eu/system/files/2023-07/Adequacy%20decision%20EU-US%20Data%20Privacy%20Framework_en.pdf (Accessed 27 March 2024).
[60] Article II (1), Annex 1 – EU-U.S. Data Privacy Framework Principles issued by the U.S. Department of Commerce.

**4.3.3 Accountability for Onward Transfer:** Organizations must adhere to the Notice and Choice Principles when transferring personal information to third-party controllers, ensuring that the data is used only for specified purposes and with the individual's consent. Contracts with these controllers must guarantee the same level of protection as outlined in the principles and require notification if this cannot be maintained, prompting appropriate action. When transferring data to third-party agents, organizations must ensure limited and specified purposes, verify privacy protection levels, and monitor compliance with principles. Agents must notify the organization of any inability to maintain protection levels, and organizations must promptly rectify any unauthorized processing. Additionally, organizations may need to provide contract privacy provisions to the Department upon request.[62]

**4.3.4 Security:** Organizations handling personal information must implement reasonable and appropriate measures to safeguard it against loss, misuse, and unauthorized access, disclosure, alteration, or destruction. These measures should consider the risks associated with processing and the type of personal data involved.[63]

**4.3.5 Data Integrity and Purpose Limitation:** Personal information should be limited to what is relevant for processing purposes and must not be processed in a manner incompatible with its original purpose. Organizations must ensure that personal data is reliable, accurate, complete, and current for its intended use, adhering to these principles for as long as they retain such information. Information identifying individuals should only be retained as long as necessary for processing purposes. However, organizations may retain data for longer periods for purposes such as archiving, journalism, research, or statistical analysis, provided they comply with other principles of the EU-U.S. DPF. Organizations should take reasonable measures to comply with this provision. In summary, personal data should be relevant and processed appropriately, with identifiable information retained only as long as necessary for processing purposes. Organizations may retain data for longer periods for specific purposes but must ensure compliance with relevant principles.[64]

**4.3.6 Access:** Individuals should have access to their personal information held by an organization and the ability to correct, amend, or delete inaccurate data or data processed in violation of the principles. However, access may be denied if providing it would be disproportionate to the privacy risks or would violate the rights of others.[65]

---

[61] Article II (2), Annex 1 – EU-U.S. Data Privacy Framework Principles issued by the U.S. Department of Commerce.
[62] Article II (3), Annex 1 – EU-U.S. Data Privacy Framework Principles issued by the U.S. Department of Commerce.
[63] Article II (4), Annex 1 – EU-U.S. Data Privacy Framework Principles issued by the U.S. Department of Commerce.
[64] Article II (5), Annex 1 – EU-U.S. Data Privacy Framework Principles issued by the U.S. Department of Commerce.

**4.3.7 Recourse Enforcement and Liability:** Effective privacy protection requires robust mechanisms to ensure compliance with the principles, provide recourse for individuals affected by non-compliance, and impose consequences on organizations that fail to follow the principles. These mechanisms include independent recourse mechanisms for resolving complaints, verification of privacy practices, and obligations to remedy non-compliance. Organizations must promptly respond to inquiries and complaints, cooperate with DPAs, and arbitrate claims according to specified procedures. In cases of onward transfer, organizations remain liable for the processing of personal information by their agents. Court orders or directives from U.S. statutory bodies regarding non-compliance must be reported publicly, and the Department serves as a point of contact for DPAs to address compliance issues. The FTC and DOT prioritize referrals of non-compliance and exchange information with EU authorities, subject to confidentiality requirements.[66]

## 5 Opportunities and Challenges of the EU-U.S. DPF for the Healthcare Sector

While the EU-U.S. DPF is a welcome initiative, its viability in the healthcare sector is yet to be determined. It certainly offers promising potential to harmonize some of the legal standards and facilitate smoother data exchange. Under this framework, research and health care institutions could potentially collaborate with EU counterparts with less hindrance, benefiting from a shared understanding of data protection principles. This alignment not only simplifies data exchange but may also foster trust among participating organizations and individuals. Even for entities not currently involved in receiving EU patient data, this new framework provides the potential to enhance organizational privacy practices, crucial in today's interconnected healthcare landscape.[67]

However, as with any regulatory framework, there are challenges to consider. Despite its potential benefits, several challenges may impede the effective use of the framework for health data transfer. One significant obstacle arises from the divergence in privacy regulations between the U.S. and the EU. While the US mandates compliance with the Health Insurance Portability and Accountability Act (HIPAA)[68] for covered entities, a comprehensive federal data protection law is lacking, resulting in varying state-level regulations. In contrast, the GDPR imposes stringent requirements on all EU member states and organizations handling EU data.[69]

---

[65] Article II (7), Annex 1 – EU-U.S. Data Privacy Framework Principles issued by the U.S. Department of Commerce.
[66] Article II (8), Annex 1 – EU-U.S. Data Privacy Framework Principles issued by the U.S. Department of Commerce.
[67] Tschider, Corrales Compagnucci and Minssen (2024).
[68] Health Insurance Portability and Accountability Act (HIPAA); Kennedy–Kassebaum Act, or Kassebaum–Kennedy Act).
[69] Tschider, Corrales Compagnucci and Minssen (2024).

In April 2024, two members of the U.S. Congress introduced a bipartisan federal privacy bill, known as the American Privacy Rights Act. This proposed legislation aims to establish a comprehensive national standard for data privacy and security, encompassing measures such as data minimization and consumer rights regarding targeted advertising. The bill grants consumers the right to access, correct, export, or delete their data. Additionally, it includes provisions for data security, and the establishment of a national data broker registry. It also prohibits organizations from enforcing mandatory arbitration in cases of significant privacy harm. In a section addressing civil rights, the bill prohibits companies from using personal information to discriminate against individuals. It empowers individuals to opt out of a company's use of algorithms in decisions regarding healthcare and other areas.[70]

Unlike previous attempts, this bill seeks to preempt state privacy laws and includes a private right of action. However, the emergence of this bill amidst a backdrop of increasing state-level privacy legislation may pose challenges due to regulatory divergence. Nonetheless, stakeholders view this proposal as a crucial step towards addressing the urgent need for privacy reform and ensuring robust data protection for all individuals in the digital age.[71]

Another step in overcoming these obstacles involves aligning the interpretation of the EU-U.S. DPF with forthcoming EU regulations, such as the European Health Data Space (EHDS).[72] The EHDS holds significant potential in addressing these hurdles. Functioning as a secure and standardized platform for health data sharing, the EHDS aims to maintain GDPR compliance while facilitating smooth data exchange for healthcare research and advancement. It aims to establish consistent regulations and guidelines for health data usage, thereby reducing legal uncertainties. The EHDS endeavors to promote cross-border health data exchange, enhancing the quality of healthcare, research initiatives, and policy development, streamlining processes, and improving efficiency.[73]

Navigating through these diverse regulatory frameworks can pose challenges. Compliance with the EU-U.S. DPF needs adherence to principles aimed at safeguarding patient and research participant privacy. These principles include limitations on the collection and use of personal information, explicit consent requirements, and provisions for data subject rights. Health professionals and organizations must adeptly navigate these new regulatory approaches to ensure seamless data transfer while maintaining compliance with both EU and U.S. laws.[74]

Implementing the EU-U.S. DPF effectively requires the adoption of robust data security measures to prevent unauthorized access and breaches. While the framework offers broad guidance on security requirements, organizations can benefit from insights gleaned from established standards like HIPAA's Security Rule[75] and international frameworks such as the NIST Cybersecurity Framework.[76] Prioritizing data security instills confidence among patients and research participants in the protection of their sensitive information. This entails employing encryption, pseudonymization, and anonymization techniques to safeguard health data during both transmission and storage.[77]

---

[70] Bracy (2024).
[71] Bracy (2024).
[72] Proposal for a regulation – The European Health Data Space (EHDS), available at: https://health.ec.europa.eu/publications/proposal-regulation-european-health-data-space_en (Accessed 27 March 2024).
[73] Legio-Quigley et al. (2024).
[74] Tschider, Corrales Compagnucci and Minssen (2024).
[75] The HIPAA Security Rule mandates that physicians safeguard patients' electronically stored protected health information (ePHI) by implementing suitable administrative, physical, and technical safeguards. These measures are essential to uphold the confidentiality, integrity, and security of this information. See HIPPA Security Rules & Risk Analysis, available at: https://www.ama-assn.org/practice-management/hipaa/hipaa-security-rule-risk-analysis (Accessed 15 April 2024).

Emerging technologies like homomorphic encryption[78] provide additional layers of security, enabling the sharing and analysis of health data without compromising its integrity. By combining secure multiparty computation with homomorphic encryption can offer a powerful solution, capable of delivering scalable and robust data protection measures in distributed data sharing environments, such as those found in health research and clinical trials. This capability directly tackles vulnerabilities inherent in traditional data protection methods, opening avenues for secondary use and research, particularly in sensitive fields like healthcare.[79] As such, these advancements pave the way for innovative applications that prioritize both data security and the advancement of medical knowledge.

Furthermore, the question arises whether the EU-U.S. DPF can effectively accommodate recent technological advancements like generative AI, which has the potential to enhance the quality of experiences in medicine and healthcare. This includes the development of AI-powered conversational user interfaces, leading to more user-friendly interactions. As generative AI continues to progress and tailor its capabilities to the specific needs of the medical field, it is poised to play an increasingly integral role to healthcare delivery. Additionally, it can facilitate patient interactions by establishing dynamic informed consent protocols.[80] Moreover, adhering to data minimization principles, which advocate for the collection and processing of only essential data, assists organizations in meeting GDPR compliance requirements.[81]

Finally, while the EU-U.S. DPF carries high hopes, its resilience in a potential *Schrems III* case remains uncertain. Despite representing an advancement over the invalidated Privacy Shield, doubts linger about the framework's ability to withstand scrutiny by the CJEU. Concerns persist regarding the sufficiency of safeguards for U.S. signals intelligence activities, which may not align with the rigorous requirements of the GDPR and the EU CFR.[82]

Before the approval of the framework, the European Parliament raised concerns about its adequacy, anticipating legal challenges. DPAs such as the Baden Wuerttemberg DPA in Germany, have also raised concerns about the adequacy of the safeguards provided by the EO in meeting CJEU standards. It questioned the effectiveness of an EO in implementing GDPR safeguards without U.S. legislative approval, and casted doubts about the enforceability of EO compliance by EU citizens.[83]

Regarding the new safeguards concerning signals intelligence activities, the German DPA sought clarification on how the EO interacts with other established U.S. regulations, such as the CLOUD Act[84] and challenged the interpretation of the proportionality principle in data collection. It criticized the bureaucratic requirements for EU citizens to file complaints with the CLPO and the lack of transparency in informing complainants about potential intelligence activities by U.S. authorities. Furthermore, the German DPA expressed concerns regarding the independence of the newly established DPRC, which falls under the authority of the Attorney General within the executive branch. This setup diverges from the traditional concept of judicial independence observed in courts established within the judicial branch. Considering these issues, the German DPA questioned the European Commission's ability to assess the adequacy of U.S. data protection solely based on the EO.[85]

---

[76] See, The NIST Cybersecurity Framework (CSF) 2.0, available at: https://nvlpubs.nist.gov/nistpubs/CSWP/NIST.CSWP.29.pdf (Accessed 15 April 2024).
[77] Tschider, Corrales Compagnucci and Minssen (2024).
[78] Corrales Compagnucci et al. (2019), pp. 144-155.
[79] Corrales Compagnucci et al. (2024), pp. 168-169.
[80] Zhang and Kamel Boulos (2023), p. 286.
[81] Legio-Quigley et al. (2024).
[82] See, e.g., Deslorieux (2024).
[83] Hanssen (2022).

Shortly after the decision, privacy advocates, including non-profit organization Noyb led by Max Schrems announced their intentions to challenge it. According to Noyb, the framework is largely a replication of the Privacy Shield. Despite efforts to present it as new, little has changed in U.S. law, notably regarding FISA 702's deficiencies. The "magic tricks" of renaming and slight modifications fail to address fundamental privacy concerns, setting the stage for another potential CJEU invalidation.[86]

French Parliament member Philippe Latombe who also serves on the French DPA (CNIL), filed two challenges before the EU General Court:[87] one to immediately suspend the agreement, and another regarding the agreement's content.[88] The request, spanning 33 pages with annexes, invoked Article 263 of the Treaty on the Functioning of the European Union (TFEU) regarding regulatory acts directly affecting individuals. Latombe's challenges, criticized the framework for its failure to adequately protect private and family life, lack of compliance with GDPR, and linguistic shortcomings. He presented the following primary legal arguments:[89]

1. Lack of effective remedy: Latombe's request criticized the absence of assurances regarding the right to an effective remedy, specifically highlighting the lack of transparency in the newly established DPRC procedure.
2. Breach of minimization and proportionality principles: Latombe also contended that the EU-U.S. DPF violates the minimization and proportionality principles of the GDPR. He argues that the alleged "bulk collection of personal data" by U.S. surveillance authorities contravenes these principles.
3. Language accessibility: Latombe emphasized the need for the EU-U.S. DPF decision to be available in the official languages of the EU, as it was only accessible in English.

---

[84] The Clarifying Lawful Overseas Use of Data Act (CLOUD Act) is a United States federal law enacted in 2018 by the passing of the Consolidated Appropriations Act PL 115–141, Division V. It primarily amends the Stored Communications Act (SCA) of 1986 to empower federal law enforcement to compel U.S.-based technology companies, through warrant or subpoena, to provide requested data stored on servers, regardless of whether the data are stored in the U.S. or abroad.
[85] Hanssen (2022).
[86] Noyb, European Commission gives EU-U.S. data transfers third round at CJEU (10 July 2023), available at: https://noyb.eu/en/european-commission-gives-eu-us-data-transfers-third-round-cjeu (Accessed 15 April 2024).
[87] Latombe filed the challenges as a private citizen of the Union and not in his professional capacities.
[88] Kayali (2023).
[89] Navarro and Schwartz (2023).

In a recent legal development on 12 October 2023, the EU General Court has denied his request for interim measures to halt the EU-U.S. DPF. The Court deemed Latombe unable to demonstrate the presence of specific serious and irreparable harm caused by the framework, either individually or collectively, a prerequisite for urgent interim measures. Latombe's application lacked specificity regarding the personal harm he would suffer from using the EU-U.S. DPF mechanism. While he highlighted general concerns about the framework, he did not clarify the nature of the damage he would personally face. The Court emphasized that the party seeking interim measures must demonstrate an imminent risk of severe, irreparable harm. Mere prima facie evidence, even if substantial, cannot suffice without such a demonstration, except in limited circumstances. Hence, the Court rejected Latombe's request for interim relief without delving into its admissibility, examining the case's merits, or balancing interests.[90]

As of the time of writing this article, Latombe has lodged an appeal challenging the Court's dismissal of his initial petition aiming to annul the framework. Among Latombe's arguments in the appeal is the contention that the U.S. DPRC lacks independence, as its creation originates from a presidential EO rather than congressional legislation. Additionally, Latombe highlights the deficiency in U.S. law regarding comprehensive safeguards against automated decision-making, along with the issues surrounding the principles of necessity, proportionality, and the right to effective judicial protection.[91]

# 6 Conclusion

This article explored the EU-U.S. DPF and its impact on health data sharing across the Atlantic, outlining its benefits and challenges. It clarified the framework's requirements and offered guidance for healthcare organizations to navigate its implementation and compliance. While the framework offers advantages like legal compliance and streamlined organization relations, it also poses challenges such as regulatory enforcement, due diligence requirements, and legal ambiguity. Thus, healthcare organizations must carefully weigh these factors before engaging in the framework.

Drawing on the analogy of the Ouroboros introduced earlier, the article questions whether the European Commission is metaphorically "eating its own tail" by grappling with legal challenges that scrutinize the effectiveness of its privacy protection mechanisms. Just as the Ouroboros symbolizes the interplay of opposing forces, the European Commission must delicately balance privacy rights with facilitating international data transfers in an interconnected world.

Ultimately, the Ouroboros narrative serves as a poignant reminder of the cyclical nature of challenges and underscores the necessity for continual adaptation in privacy and data protection law. As the European Commission strives for effective data transfer mechanisms, it must break free from this cycle and forge new paths toward enhanced privacy and data protection across the Atlantic.

---

[90] Saras and Pridas (2023).
[91] *Latombe v European Commission* (Case T-553/23).

**Acknowledgement:** This research was funded by a Novo Nordisk Foundation grant for a scientifically independent Inter-CeBIL program (grant agreement number NNF23SA0087056). The opinions expressed are the author's own and not of his respective affiliations. The author declares no conflicts of interests.